%% file: ACM-PHYSICS-FROM.tex
\documentclass[sigconf]{acmart} % submission format
\setcopyright{none} % only in review format

\usepackage[super]{nth}
\usepackage{algorithm}
\usepackage{algpseudocode}
\usepackage{amsmath}
\newcommand{\norm}[1]{\lvert #1 \rvert}
\AtBeginDocument{%
  \providecommand\BibTeX{{%
    \normalfont B\kern-0.5em{\scshape i\kern-0.25em b}\kern-0.8em\TeX}}}

\def\figscale{.4} % use .5 for the two-column submission
\begin{document}
% \title{Design Algorithm and Fabrication of a Generative Physics-based Double-curvature Structure with Planar Hexagonal Panels}
\title{Analysis of Design Algorithms and Fabrication of a Graph-based Double-curvature Structure with Planar Hexagonal Panels}
\author{Mehdi Gorjian}
\affiliation{
  \institution{Texas A\&M University}
  \city{College Station}
  \state{Texas}
  \country{USA}
  }
\email{mgorjian@tamu.edu}

\author{Gregory A. Luhan}
\affiliation{
  \institution{Texas A\&M University}
  \city{College Station}
  \state{Texas}
  \country{USA}
  }
\email{gregory.luhan@tamu.edu}

\author{Stephen M. Caffey}
\affiliation{
  \institution{Texas A\&M University}
  \city{College Station}
  \state{Texas}
  \country{USA}
  }
\email{stephencaffey@tamu.edu}
\input{data/01-abstract}
\input{ccsxml}




\begin{teaserfigure}
  \centering
  \includegraphics[width=.6\textwidth]{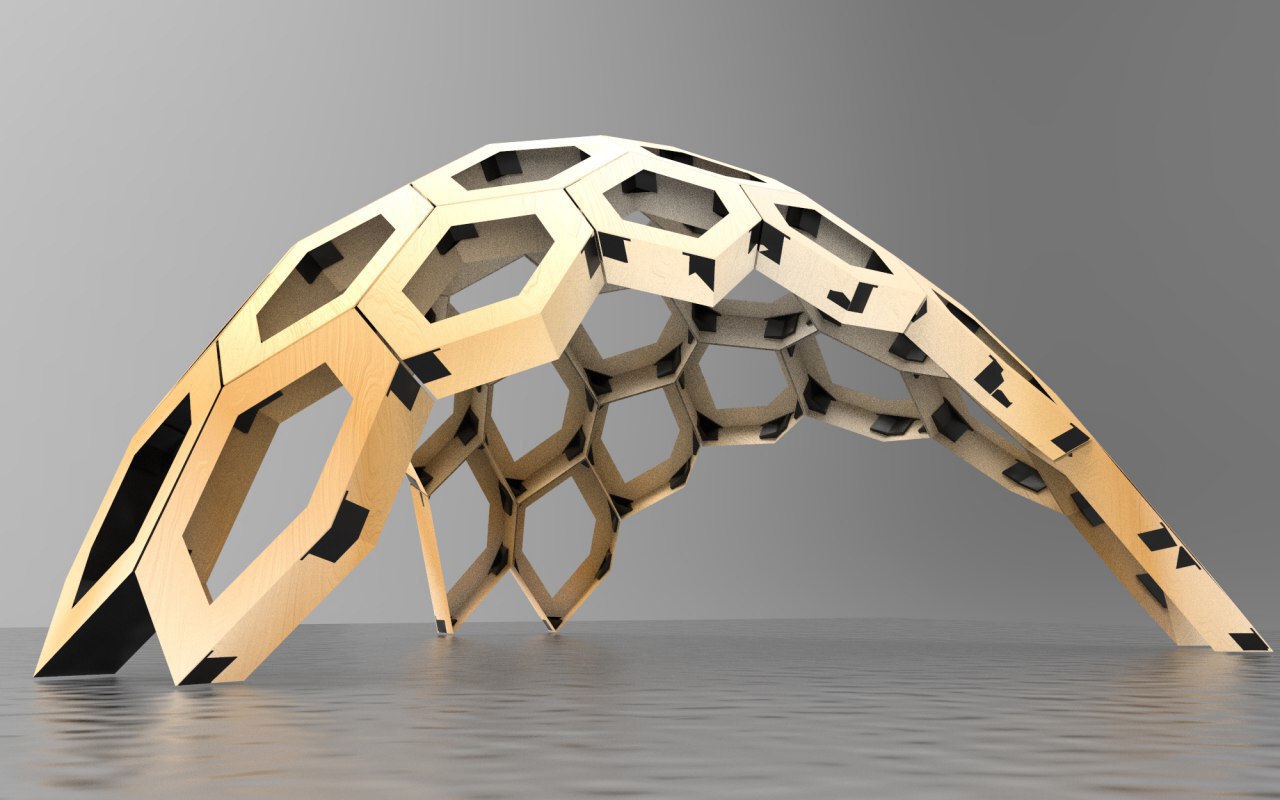}
  \caption{Double-curvature Fabricated Pavilion}
  \Description{}
  \label{fig:pavilion-teaser}
\end{teaserfigure}


\maketitle
%%%%%%%%%%%%%%%%%%%%%%%%%%%%%%%%%%%%%%%%%%%% PAPER BODY
\input{data/02-introduction}
\input{data/03-related-work}
\input{data/04-method}
\input{data/05-experimental-result}

\input{data/06-conclusion}
\bibliographystyle{ACM-Reference-Format}
\bibliography{data/ACM-PHYSICS-FROM}

\end{document}

%% file: data/01-abstract.tex
\begin{abstract}
This paper presents a novel algorithmic framework for the computational design, simulation, and fabrication of a hexagonal grid-based double-curvature structure with planar hexagonal panels. The journey begins with constructing a robust data structure through the meticulous subdivision of an equilateral triangle surface, forming a foundational triangular grid. This grid is the basis for a graph that encapsulates hexagons, laying the groundwork for simulating dynamic interactions and form-finding. The developed algorithm ensures a well-structured hexagonal grid data representation, and the experimental results showcase the successful implementation of the algorithm, leading to the fabrication of planar hexagons mirroring physics-generated mesh surfaces.
\end{abstract}

%% file: ccsxml.tex

\ccsdesc[500]{General and reference~Design}

%% file: data/02-introduction.tex
\section{Introduction}
The field of digital design has witnessed significant advancements, particularly with the emergence of physics-based generative design methodologies. These innovative approaches exemplify synthesizing intricate natural phenomena within a dynamic digital environment characterized by sequential formation and transformation processes. Using physics-based simulations \cite{thakur_survey_2009} allows for the gradual development of exceptionally flexible forms, showcasing the dynamic interplay of various simulation elements. Within this realm, diverse techniques have been employed to generate forms from physics-based systems, each offering unique insights into the design process. One notable approach involves the utilization of deformation, collision, or equilibrium systems \cite{nicolis_physics_1993}, wherein the behavior of virtual materials or particles is simulated to achieve desired outcomes. This simulation-driven design \cite{karlberg_state_2013} process facilitates the exploration of a broad spectrum of complex and dynamic forms, enabling designers to harness the creative potential embedded within simulated physical interactions.

Specifically, the paper delves into analyzing design algorithms and fabricating a graph-based double-curvature structure with planar hexagonal panels. The focus lies on the intersection of physics-based generative design \cite{attar_physics-based_2009} and lightweight structures \cite{schlaich_lightweight_2000}, a classic challenge in structural form-finding. Notably, lightweight structures present a unique set of design considerations, necessitating innovative solutions to achieve both structural integrity and aesthetic appeal. Various techniques have been explored in the pursuit of form-finding for lightweight structures, including dynamic equilibrium and force-density methods \cite{schek_force_1974}. These approaches are instrumental in creating grid-like structures, offering insights into optimizing structural elements to achieve a delicate balance between material efficiency and structural stability. The integration of dynamic equilibrium principles allows for the exploration of equilibrium states under varying loads, providing a foundation for the development of structurally efficient configurations.

Furthermore, the consideration of force density within the form-finding process underscores the significance of understanding the distribution of forces within the structure. This knowledge proves crucial in achieving optimal load paths and, consequently, an efficient use of materials. By leveraging force density techniques, designers can iteratively refine the geometry of lightweight structures, ensuring they meet both functional and aesthetic requirements. The paper sets the stage for a comprehensive exploration of the symbiotic relationship between physics-based generative design, lightweight structures, and advanced form-finding techniques. % Through a nuanced analysis of design algorithms and the fabrication process, the research contributes valuable insights to the broader discourse on digital design, providing a platform for the development of innovative structures that seamlessly blend functionality and aesthetics in the digital era.

%% file: data/03-related-work.tex
\section{Related Work}
\subsection{Physics Generated Systems}
\textit{Dynamic equilibrium} approaches, including the \textit{Dynamic Relaxation} and \textit{Particle Spring Systems}, transform the form-finding challenge into a dynamic problem. They simulate the behavior of tangible form-finding research, such as a network of strings suspended from the ceiling or a soap bubble. These techniques employ a system of linear components with simulated flexible behavior, initial length, and aggregated node masses. The model oscillates around the point of equilibrium position until it reaches an absolute equilibrium state. Within this repetitive strategy, the location and velocity of nodes in networks are updated since all of the forces on a node are in equilibrium. \textit{The Dynamic Relaxation Method (DRM)} is one of the suitable numerical procedures for nonlinear structural analysis \cite{rezaiee-pajand_fast_2019}. A system of linear components with simulated flexible behavior, starting length considerations, and aggregated node masses is used in the application of dynamic equilibrium techniques. This computational approach captures the subtle interactions that result in complex structures by simulating the behavior of real materials. Until it converges to an absolute equilibrium state, the model created using these procedures oscillates about the equilibrium position. The position and velocity of nodes within the simulated network are continuously updated throughout this iterative process to guarantee that all forces acting on a node are in an equilibrium condition. DRM is also used for calculating and simulating the Active Bending Structures. Active Bending is an increasingly popular construction technique that uses elastically bent structural members to form complex curved shapes \cite{rombouts_fast_2019}. This technique not only showcases the adaptability of DRM but also highlights the integration of computational methodologies in pushing the boundaries of construction possibilities. DRM's efficacy in handling nonlinearities makes it particularly well-suited for exploring the behavior of complex structures under varying conditions. This versatility extends beyond traditional structural analysis, as DRM finds application in calculating and simulating Active Bending Structures. Active Bending Structures, with their innovative use of materials and dynamic design principles, exemplify the synergy between computational simulations and architectural creativity. The increasing popularity of Active Bending Structures underscores a broader trend in contemporary construction methodologies, where computational tools play a pivotal role in shaping and realizing architectural visions. As these dynamic equilibrium approaches continue to evolve and find broader applications, they contribute not only to the theoretical underpinnings of form-finding but also to the practical implementation of cutting-edge and aesthetically compelling structures in the built environment.

\textit{The Force Density Method (FDM)} is based upon the force-length ratios or force densities, which are defined for each branch of the net structure \cite{schek_force_1974}. In general, these solution approaches help to locate the state of equilibrium of a structure's network with an appropriate amount of internal force, such as membrane structures. As Kilian and Ochsendorf stated, ``In the early \nth{20} century, Antoní Gaudi employed hanging models in the form-finding processes for the chapel of the Colonia Guell and the arches of the Casa Mila. Such hanging forms are often called \textit{funicular}, derived from the Latin word \textit{funiculus}, meaning thin cord or rope because they represent the shape taken by a thin cord acting in pure tension under a given set of loads'' \cite{kilian_particle-spring_2005}. The Force Density Method operates by assigning force densities to different segments of the structural network, providing a means to distribute forces in a manner that achieves equilibrium. This approach is particularly advantageous in the analysis of membrane structures, where the interplay of forces is critical to maintaining stability. By employing force-length ratios, the FDM aids in locating the state of equilibrium for complex structures, offering a valuable computational tool for architects and engineers engaged in form-finding processes. The reference to Gaudi's use of hanging models highlights a longstanding tradition of employing physical prototypes to inform architectural design. This connection between historical practices and contemporary computational methodologies underscores the timeless relevance of form-finding principles. As computational tools continue to advance, the Force Density Method stands out as a versatile and effective approach, bridging the historical roots of form-finding with the computational sophistication of the present day.

\noindent\textit{Particle spring systems} are based on lumped masses, called particles, which are connected by linear elastic springs. Each spring is assigned a constant axial stiffness, an initial length, and a damping coefficient. Springs generate a force when displaced from their rest length. External forces can be applied to the particles, as in the case of gravitational acceleration \cite{kilian_particle-spring_2005}. Runge-Kutta algorithm \cite{baraff_large_1998} was used by Kilian and Ochsendorf for designing particle spring systems.

\subsection{Advancements in Planarization Methods}
Cutler and Whiting \cite{cutler_constrained_2007} computed planar panels for a given surface using the variational surface approximating method \cite{cohen-steiner_variational_2004}. However, the resulting planar polygonal meshes, characterized by valence-3 vertices, lacked the desired control over face and side counts. Recognizing the need for more precise and versatile planarization techniques, subsequent research has delved into refining and expanding upon these foundational approaches. Pottmann et al. \cite{pottmann_geometry_2007} introduced a sophisticated method for computing the two sides of a Koebe polyhedron based on planar polygonal geometry. Nevertheless, their form--finding approach is limited and fails to produce results for general shapes. Despite their contribution, the resulting meshes were prone to self-intersecting features, compromising aesthetic appeal and functional integrity. Almegaard et al. \cite{almegaard_surfaces_2007} determined a piecewise linear supportive function of a surface to generate a planar hexagonal geometry using duo projection. These outcomes might include self--intersecting features that detract from the mesh's aesthetics and functionality. Alexa and Wardetzky \cite{alexa_discrete_2011} demonstrate building a Laplacian operator for non-triangular geometry. As in the triangular case, their operator can smooth a mesh using Laplacian smoothing, comparable to mean-curvature flow. Given a general non-polyhedral 3D mesh, discover the nearest planar polyhedral mesh with an identical or comparable topology. Poranne et al. \cite{poranne_interactive_2013} gave a solution to this issue that facilitates a highly customizable interactive system for the designer. Hoffmann \cite{hoffmann_local_2010} demonstrates that an individual quad can be deformed despite affecting its immediate neighbors or exceeding the planarity constraints. However, it is important to note that this technique, while demonstrating localized deformations, falls short of providing a generalized solution for the deformation of entire quadrilateral meshes. % From addressing the limitations of early approaches to providing interactive tools for designers, these advancements pave the way for more precise, versatile, and user-friendly applications in the field of digital design and computational geometry. As researchers continue to innovate, the future promises even more significant strides in planarization methods, offering enhanced solutions for complex shapes and facilitating a seamless integration of computational techniques into the design process.

%% file: data/04-method.tex
\section{Design Algorithm Method}
Our comprehensive design algorithm methodology involves a structured series of steps to seamlessly integrate computational processes with geometric principles to generate a functional and aesthetically pleasing hexagonal structure. Each step contributes to the overall evolution of the design, incorporating considerations for data structure configuration, particle spring systems, planarization algorithms, and final fabrication details. Our approach to the design algorithm comprises the following steps:

\begin{enumerate}
    \item \textbf{Data Structure Configuration}
        \begin{enumerate}
            \item \textbf{Developing a Hash Map:} Establishing a robust data structure is paramount. We create a Hash Map with the face index as the key and an array of corresponding vertex indices as the value. This ensures efficient organization and retrieval of essential geometric information.
            % \item \textit{Array of vertices coordinates}
            \item \textbf{Constructing Graph of Hexagons:} To facilitate dynamic form-finding, we construct a graph of hexagons. This involves several sub-steps:
                \begin{enumerate}
                    \item \textit{Determining a Planar Surface:} Identifying a planar surface based on a triangle with equal sides, providing a foundational geometric reference.
                    \item \textit{Subdividing the Area:} Employing an equilateral triangle grid to subdivide the identified planar surface, establishing a framework for subsequent hexagonal constructions.
                    \item \textit{Undirected Cyclic Graph Construction:} Forming an Undirected Cyclic Graph from the triangle grid, where each hexagon represents a node, and edges signify shared edges between hexagons.
                \end{enumerate}
        \end{enumerate}
    
    \item \textbf{Creating Particle Spring System from Graph}
        \begin{enumerate}
            \item \textbf{Employing the Graph:} Utilizing the constructed graph to define a particle spring system, establishing dynamic connections between hexagons. This step sets the stage for the simulation of structural behaviors and interactions.
        \end{enumerate}

    \item \textbf{Planarization Algorithm on Non-planar Hexagons}
        \begin{enumerate}
            \item \textbf{Performing Planarization Algorithm:} Addressing non-planarity issues, we implement a planarization algorithm on the graph. This iterative process modifies node positions until all hexagons' edges lie on a single plane, enhancing structural coherence and visual uniformity.
        \end{enumerate}

    \item \textbf{Adding Fabrication Details to the Final Model}
        \begin{enumerate}
            \item \textbf{Constructing Vertical Panels:} Establishing fabrication details, we create vertical panels across every hexagon's edge at each calculated point, ensuring both structural integrity and aesthetic coherence.
            \item \textbf{Creating 3D-Printed Junctions:} For practical assembly, we specify each corner of the hexagonal box as a proportion of each edge, facilitating the creation of 3D-printed junctions.
            \item \textbf{Adding Thickness and Fastener Holes:} Finalizing the design, we add thickness to the entire system for durability and delineate bolt and fastener holes on hexagonal frames and joints, streamlining the fabrication process.
        \end{enumerate}
\end{enumerate}

\begin{figure}
    \centering
    \includegraphics[width=\figscale\linewidth]{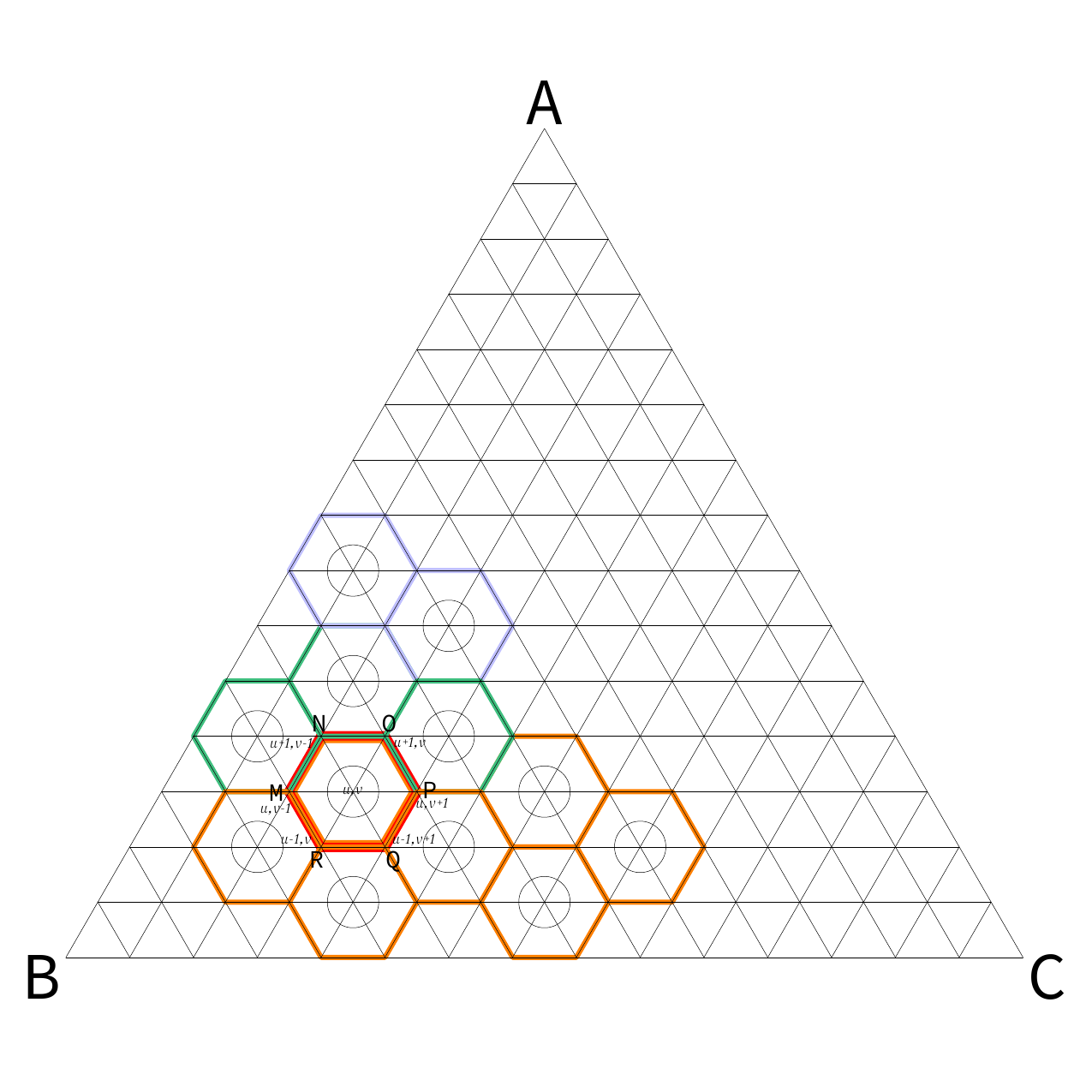}
    \caption{Adding hexagonal panels to a triangular gird}
    \label{fig:tri-grid}
\end{figure}

\subsection{Constructing Data Structure}
Our algorithmic journey commences with the construction of a robust data structure, leveraging an equilateral triangle surface denoted as $\triangle{ABC}$. To establish a grid framework for subsequent hexagonal network generation, each triangle edge is meticulously subdivided by an arbitrary factor of 3. This process culminates in the linkage of each point to its counterpart on the opposite edge, creating a foundational triangular grid (refer to Fig.~\ref{fig:tri-grid}). Subsequently, this triangular grid serves as the canvas for constructing a graph that encapsulates the hexagons within the triangular network. Employing a carefully devised algorithm, we identify groups of six nodes (M, N, O, P, Q, R) in close proximity. The parametrized u and v coordinates of each of these nodes are intricately related to the (u,v) node. Specifically, the parametrized coordinates for these neighboring nodes are as follows: $(u, v-1)$, $(u+1, v-1)$, $(u+1, v)$, $(u, v+1)$, $(u-1, v+1)$, $(u-1, v)$, $(u, v+1)$, $(u-1, v+1)$, and $(u-1, v)$. This relationship is visually depicted in Fig.~\ref{fig:tri-grid}.

This geometrically iterative process forms the foundation of our data structure, creating an intricate network of interconnected hexagons seamlessly embedded within the triangular grid. As we advance through subsequent stages of our algorithm, this meticulously constructed data structure becomes the backbone for simulating dynamic interactions, facilitating form-finding, and ultimately culminating in a structurally optimized and visually compelling design.

\begin{algorithm}
\caption{Constructing Data Structure}
\label{alg:hex-gird-algo}
\begin{algorithmic}
\State ConstructingDataStructure($G$, $F$, $S$, $Rows$, $Cols$) %\Comment{S: triangular surface, rows \texttt{==} cols}
% \State Initialize $G$ \Comment{Graph initialization}
\State define $faceIndx$ $\leftarrow$ $0$, $vertIndx$ $\leftarrow$ $0$
% \State $Map<int, int> FaceMap$
\If{($Rows$ \% $3$ \texttt{!=} $0$ \&\& $Cols$ \% $3$ \texttt{!=} $0$)}
    \State throw $invalid\_argument($"NOT a factor of 3"$)$
\EndIf

\For{($u \leftarrow 1$; $u < Rows$; $u\texttt{++}$)}
    \State define $v\_start$
    \If{$u$ \% $3$ \texttt{!=} $0$}
        \State $v\_start \leftarrow u$
    \Else
        \State $v\_start \leftarrow 3$
    \EndIf
    \State $Cols$ \texttt{-=} $1$ \Comment{Each upper row has one cell less}

    \For{($v \leftarrow v\_start$; $v < Cols$; $v\texttt{+=}3$)}
        \State add each \textit{vertex index} to its corresponding face map $F$

        \State add edges to the graph
        \State increase \textit{face index} by $1$
    \EndFor
\EndFor
\end{algorithmic}
\end{algorithm}

The \textit{Constructing Data Structure} algorithm is designed to meticulously generate a comprehensive hexagonal grid data structure, crucial for subsequent stages in our computational design process. The expanded algorithm ensures a factor-of-three constraint on rows and columns, providing a robust foundation for hexagonal grid construction. The algorithm initiates with the definition of essential variables, including $faceIndx$ and $vertIndx$, signifying the indices for faces and vertices, respectively. A check is incorporated to ensure that the number of rows and columns are factors of 3, essential for the algorithm's successful execution. The nested loop structure iterates through the triangular surface, creating hexagonal cells at regular intervals. The variable $v\_start$ is intelligently determined based on the divisibility of $u$ by 3, ensuring an optimal starting point for each row. The algorithm then iterates through the columns, adding vertex indices to their corresponding face map $F$ and establishing edges within the graph $G$. The face index is incremented accordingly, contributing to the seamless creation of the hexagonal grid. This comprehensive algorithm lays the groundwork for subsequent steps in our computational design process, ensuring the generation of a well-structured hexagonal grid data representation.

\begin{figure}
    \centering
    \includegraphics[width=\figscale\linewidth]{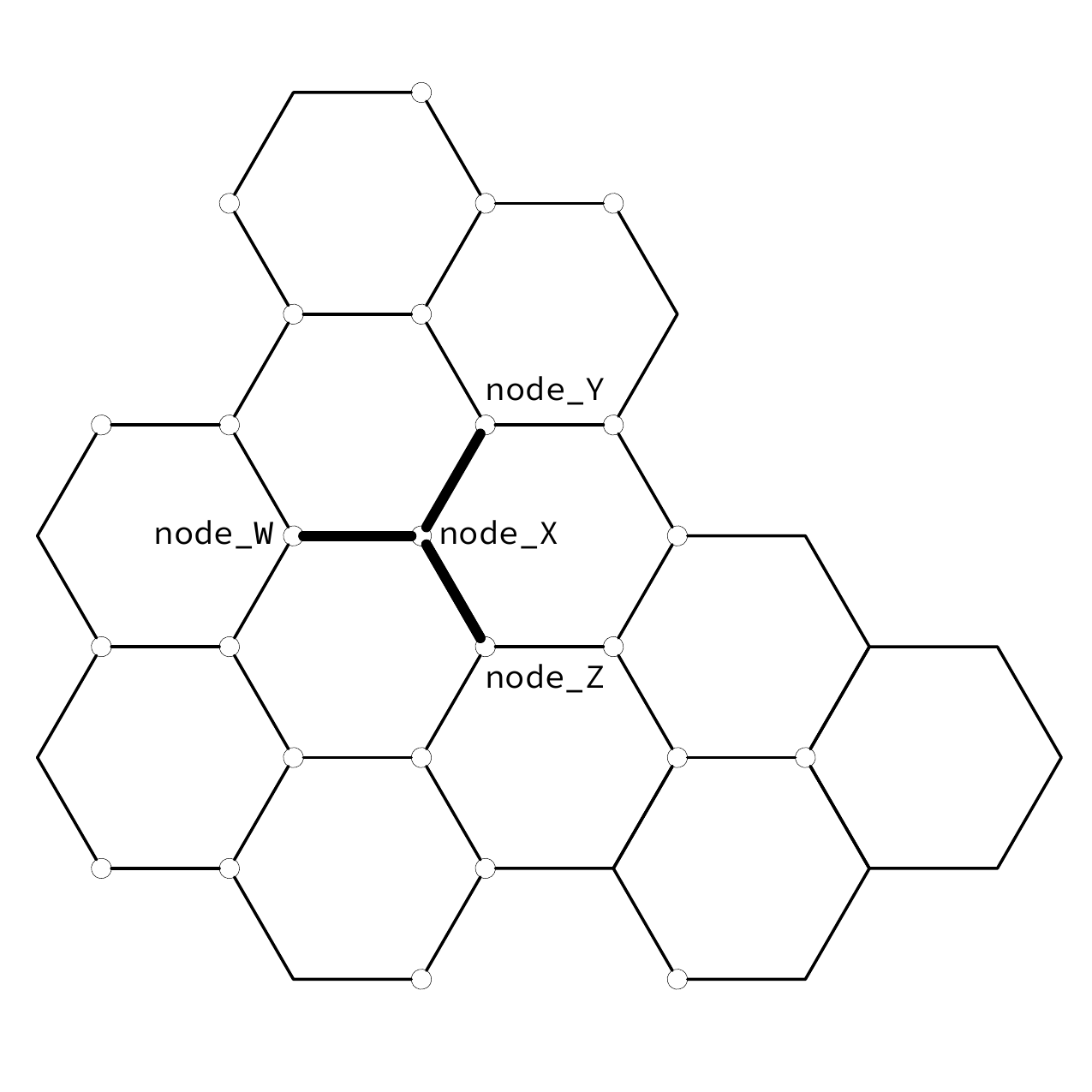}
    \caption{Constructing graph from the created hexagons}
    \label{fig:graph}
\end{figure}

The algorithm requires several crucial parameters: a Graph pointer, a Hash Map encapsulating faces with their associated vertex indices, a pointer to the base surface, and subdivision numbers. As depicted in Fig.~\ref{fig:graph}, the graph structure is intricately designed so that each node is intricately connected to three distinct nodes. The utilization of graphs contributes to creating a more robust system with significantly reduced computation time. Initially, the algorithm's complexity for constructing the graph on a triangular grid is $\mathcal{O}(n^2)$, a testament to its efficiency in handling the geometric intricacies of the design. Subsequently, the algorithm's complexity for retrieving graph members becomes linear.

The face array is pivotal in storing the node indices corresponding to each face. This array becomes instrumental in subsequent stages of the algorithm, facilitating physics simulations and expediting the planarization operation. The algorithm's efficiency lies in its ability to rapidly update the coordinates of each point within the face array during the planarization process. The robust graph structure, as illustrated in Fig.~\ref{fig:graph}, not only serves as a foundation for physics simulations but also ensures the seamless execution of the planarization operation. By leveraging the connectivity defined in the graph, the algorithm optimally handles the dynamic interactions between nodes, leading to a more accurate and efficient computational representation of the evolving hexagonal grid.

The developed algorithm's parameter requirements and the intricacies of its graph-based approach underscore its capability to handle complex geometric structures with computational efficiency. This computational framework forms the backbone for subsequent phases of the design process, contributing to the overall success of the hexagonal structure's form-finding and planarization.

\subsection{Development of Particle Spring Systems: A Multifaceted Approach}

Particle spring systems are intricately composed of \textit{Particles}, \textit{Springs}, and \textit{Forces}, each playing a pivotal role in the dynamic evolution of the system. Understanding the underlying principles of these components is essential for grasping the complexity of the algorithm.

\subsubsection{Particle Characteristics and Dynamics}

\textit{Particles}, the fundamental entities within the system, possess three defining characteristics: \textit{mass} $(m)$, \textit{position} $(x(t))$, and \textit{velocity} $(v(t))$, where velocity is the initial derivative of position, denoted as $x'(t)$. Newton's second law of motion, $F = ma$, equates the force exerted on a particle to its mass multiplied by its acceleration. Considering that $a = v'(t)$ and $v = x'(t)$, we can further simplify to $a = x''(t)$, leading to the fundamental equation of motion in the context of particle dynamics:
\begin{equation}
a = v'(t) = x''(t) = \frac{\Vec{F(t)}}{m} \label{eq:particle-acc}
\end{equation}
This equation serves as the cornerstone for understanding the acceleration of particles within the system.

\subsubsection{Spring Components: Stretch, Shear, and Bend Springs}

The structural connectivity between particles is facilitated through different types of springs: \textit{stretch springs}, \textit{shear springs}, and \textit{bending springs}.

\begin{figure}%[ht]
\centering
\includegraphics[width=\figscale\linewidth]{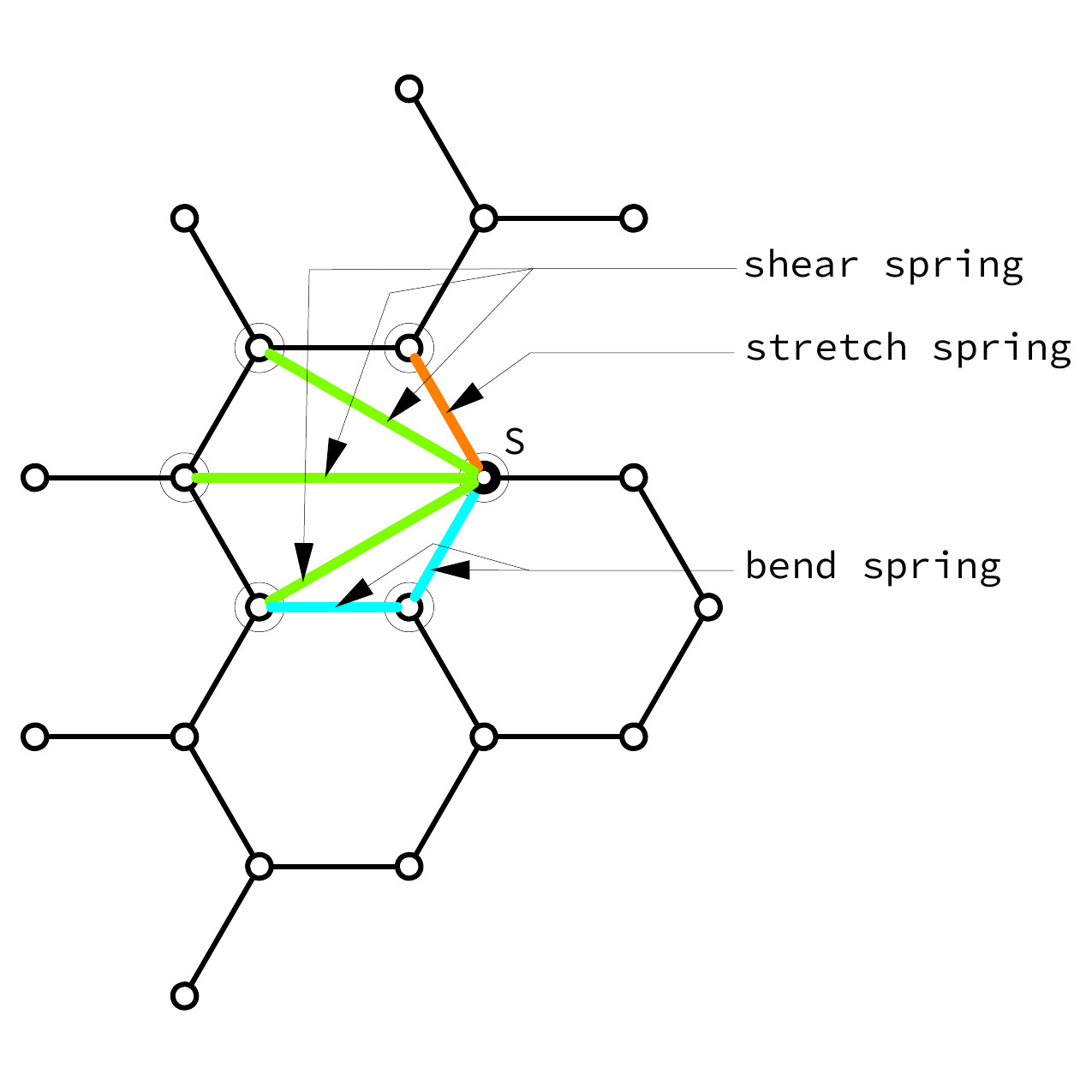}
\caption{Stretch, Shear, and Bend springs particles' links}
\label{fig:spring-system}
\end{figure}

\textit{Stretch springs} represent the structural connection between each pair of particles, creating the mesh-like framework. \textit{Shear springs} establish diagonal connections between two particles that aren't directly linked through a structural link. Finally, \textit{bending springs} form connections between a particle and the second consecutive connection in a system. The visual representation of these spring components is illustrated in Fig.~\ref{fig:spring-system}, depicting the interconnections of an arbitrary node $S$.

\subsubsection{Forces in Particle Spring Systems}

Forces are essential components that govern the behavior of particle spring systems. Four distinct types of forces contribute to the system's dynamics:

\begin{enumerate}
\item \textit{Spring forces} $(\Vec{F_{sp}})$: These forces act on the springs linking pairs of particles at positions $X_1$ and $X_2$. Governed by Hooke's Law, the spring force acting on $X_1$ is defined as:
\begin{equation}
\Vec{F_{sp}} = K(L - \norm{|X_1 - X_2|}) \frac{X_1 - X_2}{\norm{|X_1 - X_2|}} \label{eq:fsp}
\end{equation}
Here, $K$ represents the stiffness of the spring, $L$ is the rest length, and $\norm{\cdot}$ denotes the Euclidean norm.
\item \textit{Gravitational force} $(\Vec{F_{gr}})$: Modeled as a gravitational force acting on each particle, it is defined as:
\begin{equation} \label{eq:fgr}
    \Vec{F_{gr}} = (0, 0, -mg) \quad \text{where} \quad g = 9.81 \; m/s^2
\end{equation}
This force simulates the effect of gravity on the system.

\item \textit{Damping}: Damping forces can be applied to simulate a more authentic physical simulation, introducing resistance to the motion of particles.

\item \textit{Viscosity}: Similar to damping, viscosity forces can be introduced to model the effects of fluid-like interactions within the system.
\end{enumerate}
Notably, fluid forces were not applied in our algorithm, as our focus was on showcasing how external forces influence the structure's shape rather than creating a fully animated fluid simulation. Understanding the interplay of these forces is important for capturing the dynamic behavior of the particle spring system, providing a foundation for subsequent stages in the algorithm's execution. The intricate balance between these components provides a realistic and responsive simulation, enabling the algorithm to capture the hexagonal structure's evolving shape accurately.

\subsection{Panel Planarization: An Optimization Approach}
The panel planarization process, inspired by the methodology outlined in \cite{poranne_interactive_2013}, aims to achieve planarity for hexagonal faces within a mesh. The control points and solution mesh are denoted by sets $P$ and $Q$, respectively, with their individual elements represented as $p_i$ and $q_i$, where $i$ ranges from $1$ to $N_v$, the total number of vertices. The faces of the control geometry are denoted as $F$, with each face being a sorted list of vertices.

\begin{equation} \label{eq:pl}
% \begin{aligned}
    P = {\{p_i\}}_{i=1}^{N_v} \; \text{, } \;
    Q = {\{q_i\}}_{i=1}^{N_v}
% \end{aligned}
\end{equation}

The topologies of the control and solution meshes being equivalent; their face sets also share this property.

\begin{figure}
    \centering
    \includegraphics[width=\figscale\linewidth]{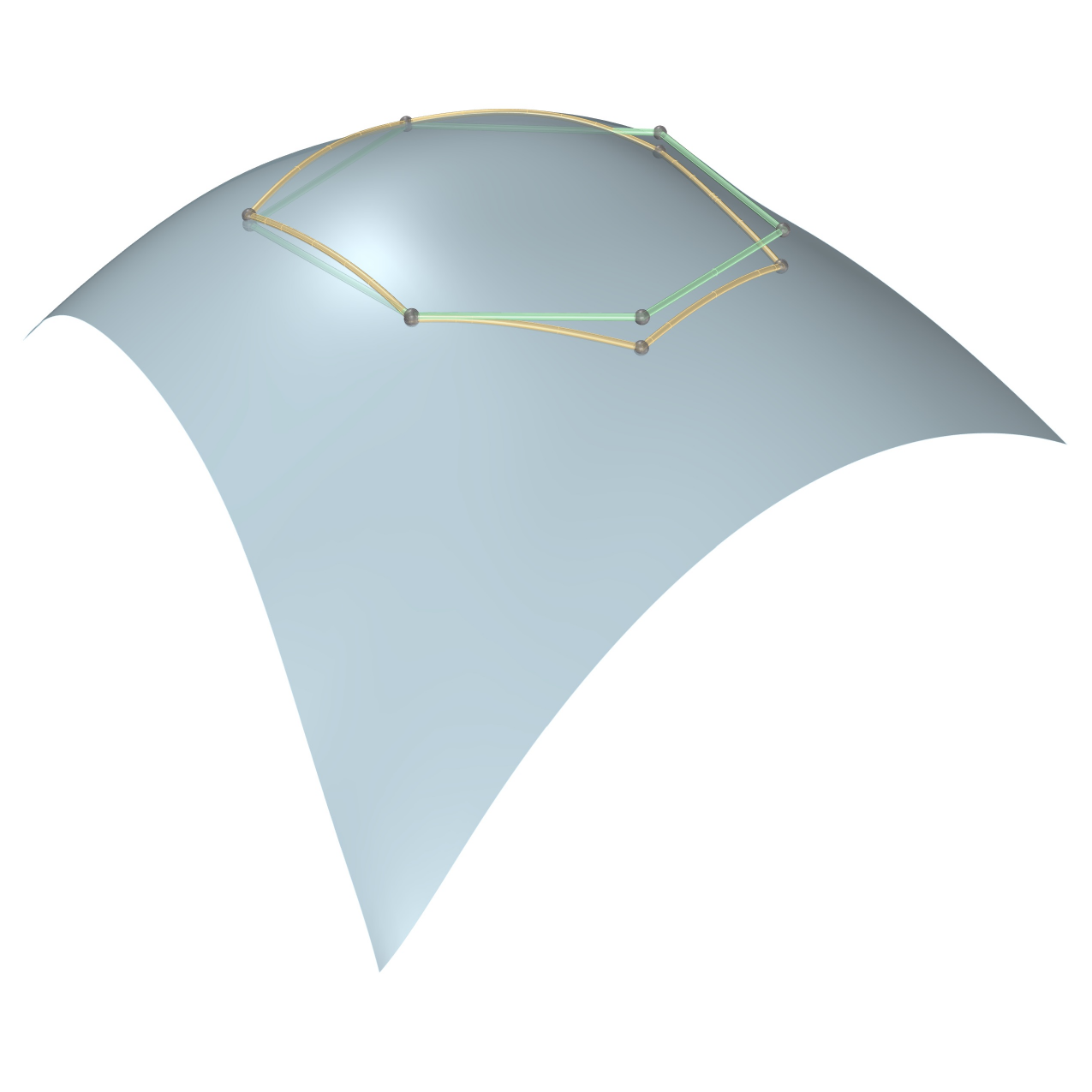}
    \caption{Planarization of an arbitrary form, where the orange polygon illustrates the control point mesh and the green polygon is the solution mesh}
    \label{fig:pl-shape}
\end{figure}

The relationship between vertices and faces is expressed through $p_i \in f_i$, signifying that vertex $p_i$ is incident to face $f_i$. The reverse notation, $f_i \in p_i$, is also employed for conciseness \cite{poranne_interactive_2013}. The set $C$ represents the corners of the mesh, defined as the set of indices of incident vertex-face pairs in the mesh, i.e., $(i,j) \in C$ if $p_i \in f_i$ (or $f_i \in p_i$).

The optimization problem is formulated based on this information, as illustrated in Fig.~\ref{fig:pl-shape}. The primary objective is to minimize the distance between each $p_i$ and $q_i$ using a least-squares approach while preserving the planarity of all $q_i$ points. Each face $f_i$ is characterized by its unit normal vector $\Vec{n_j}$ and the distance $d_j$ of its plane from the origin, forming sets $N$ and $D$ of \textit{normals} and \textit{distances}, respectively. The constraint ensures that for each face $f_i$ and each vertex $q_i \in f_i$, $q_i$ lies on the plane defined by $(n_j, d_j)$. The optimization problem is formulated as follows:

\begin{equation}
% \begin{align}
\text{\textbf{Problem:} } \min \sum_{i=1}^{N_v} ||p_i - q_i|| ^2, \;
\text{\textbf{Const.:} } n_j \cdot q_i + d_j = 0 \text{, }\forall (i,j) \in C
% \end{align}
\end{equation}

This rigorous approach seeks to achieve a balanced compromise between minimizing the distance between control and solution points and ensuring the planarity of the resulting mesh. The solution to this problem provides a geometrically optimized mesh, exemplifying the power of optimization techniques in refining complex forms.

% \noindent\textcolor{red}{MORE TO WRITE HERE...}

%% file: data/05-experimental-result.tex
\section{Experimental Results: Fabrication and Structural Reinforcement}

\begin{figure}
    \centering
    \includegraphics[width=\figscale\linewidth]{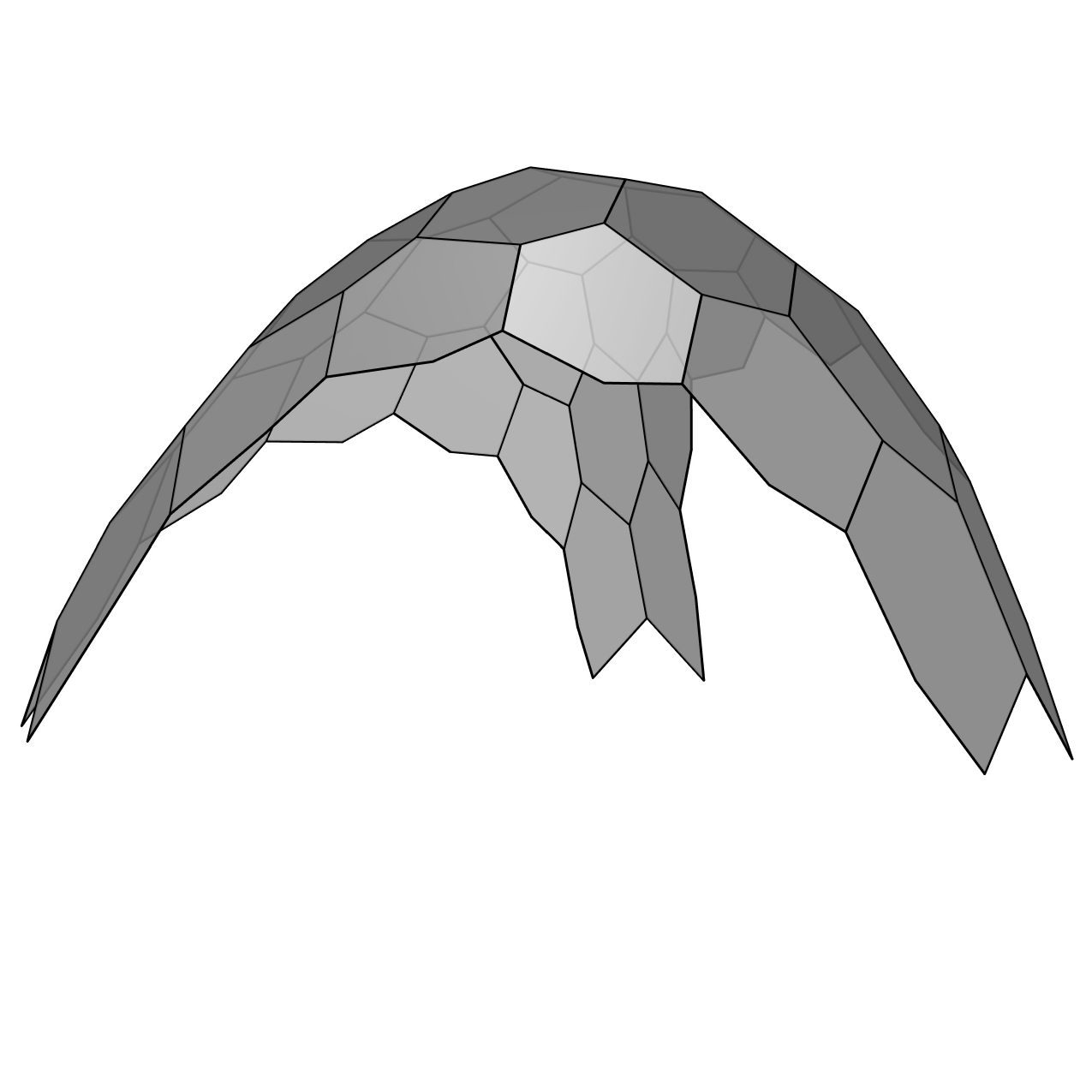}
    \caption{Hex-Planar polygons}
    \label{fig:planar-frame}
\end{figure}

The successful implementation of our algorithm has led to the production of planar hexagons closely mirroring the physics-generated mesh surface depicted in Fig.~\ref{fig:planar-frame}. Each hexagonal panel, vital for the final mesh production, was endowed with material thickness to ensure structural integrity. Building upon the hexagon polygonal boundaries, additional walls were strategically constructed.

\begin{figure}[H]
    \centering
    \includegraphics[width=\figscale\linewidth]{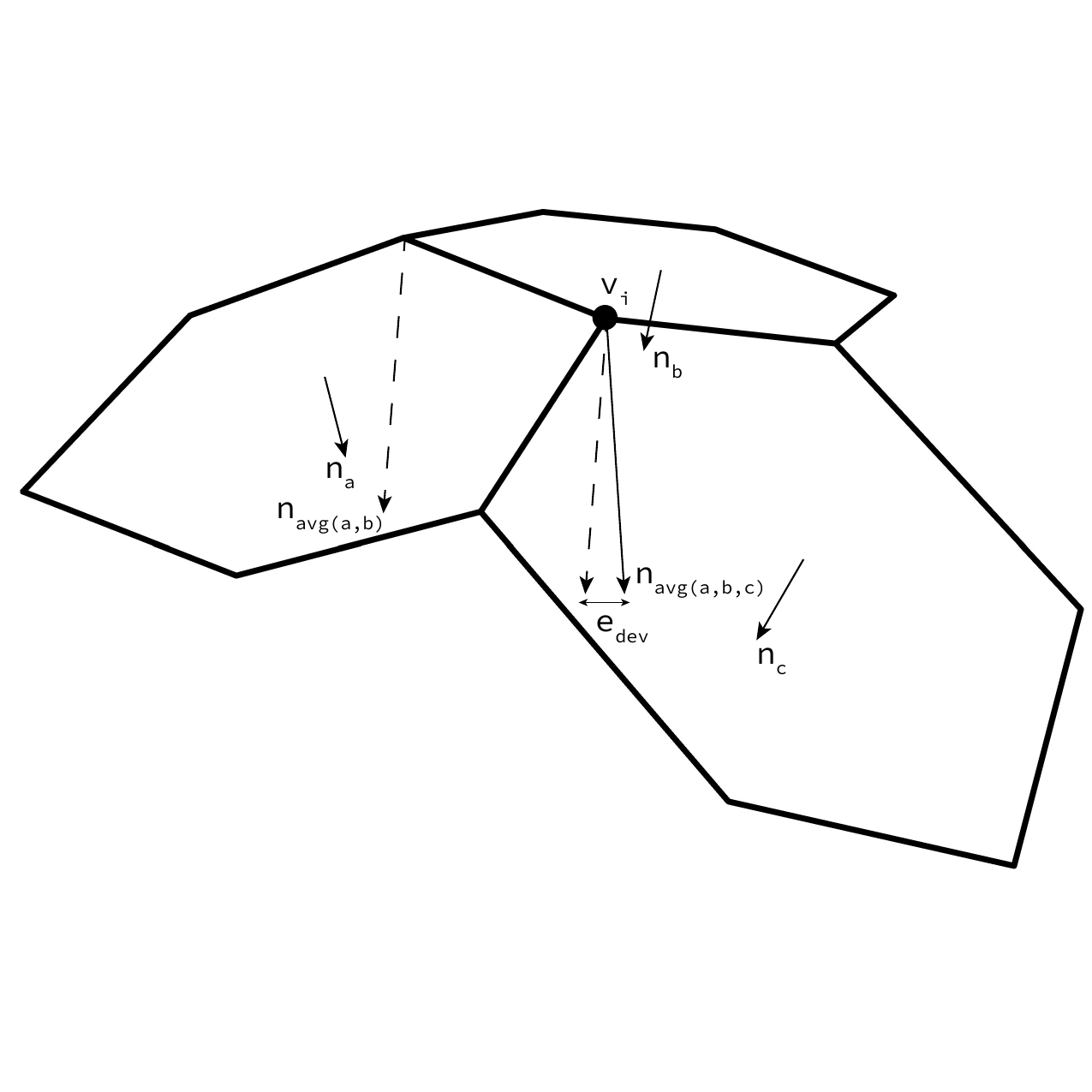}
    \caption{Maximum deviation error between the average of the three faces vs. two faces sharing the same vertex}
    \label{fig:deviation}
\end{figure}

To fortify the structure, we introduced walls along each edge. A technological challenge arose due to equipment constraints, prompting the need for a novel solution. For each vertex, we determined the average of the two ($\Vec{n_{avg(a,b)}}$) and three normal vectors ($\Vec{n_{avg(a,b,c)}}$) that shared the vertex with the corresponding edge and only vertex (Eq.~\ref{eq:avg2}), respectively illustrated in Fig.~\ref{fig:deviation}, resulting in 

\begin{equation} \label{eq:avg2}
    % n_i = \Vec{v_i} + \Vec{nt} \label{eq:vert-vec} \\
    n_{avg(a,b)} = \frac{1}{2} (n_a + n_b) , \; %\label{eq:avg2} \\
    n_{avg(a,b,c)} = \frac{1}{3} (n_a + n_b + n_c) %\label{eq:avg3}
\end{equation}
We chose to reinforce the structure by adding walls to every edge because our equipment was constrained; nevertheless; however, there was a technological issue we had to solve. The endpoints of each edge had to be translated with a vector equal to the average of the normal vectors of the three faces that shared a vertex in order to form the walls.
However, an inherent issue persisted. The translation of endpoints along the edges using the average normal vectors did not guarantee the eventual planarity of the constructed walls. To address this, we introduced a deviation error ($e_{dev}$), defined as the maximum distance between the two endpoints of $\Vec{n_{avg(a,b)}}$ and $\Vec{n_{avg(a,b,c)}}$ vectors starting at vertex $v_i$. We calculated this error for all vertices, and the deviation was consistently less than or equal to $4$mm, a reasonable threshold for a structure of this scale (Fig.~\ref{fig:deviation}).

Having successfully constructed the walls, the next step involved adding joints to hold the panel boxes together. These joints, to be manufactured using 3D printers, were constructed using the $\Vec{n_{avg(a,b,c)}}$ vector. For each edge leaving the vertex $v_i$, distinct vectors $v^1_{ij}$, $v^2_{ij}$, and $v^3_{ij}$ for face $F_j$ were calculated.

\begin{figure}[H]
    \centering
    \includegraphics[width=\figscale\linewidth]{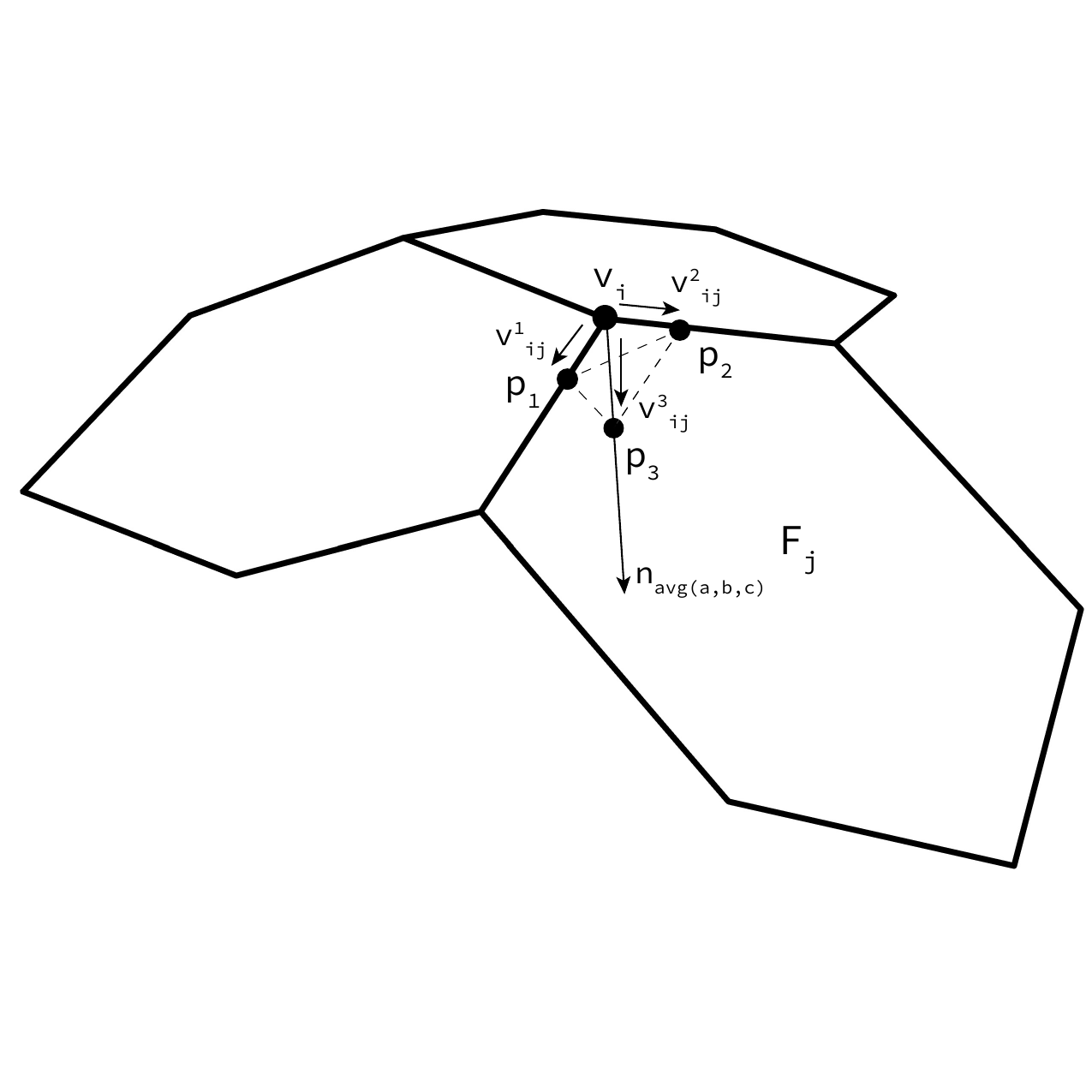}
    \caption{Joint construction}
    \label{fig:joint}
\end{figure}
\begin{figure}[H]
    \centering
    \includegraphics[width=.6\linewidth]{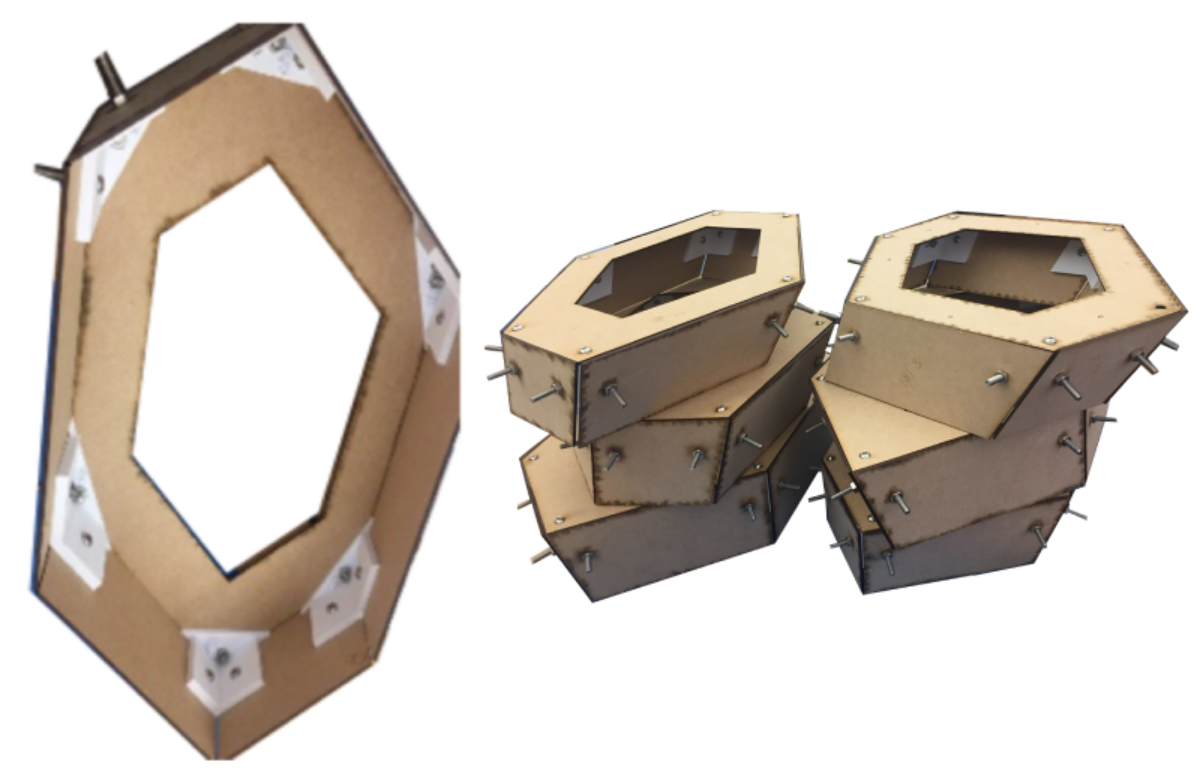}
    \caption{Fabricated joints \& panel-boxes}
    \label{fig:fabricated-panel}
\end{figure}
\begin{figure}[H]
    \centering
    \includegraphics[width=.9\linewidth]{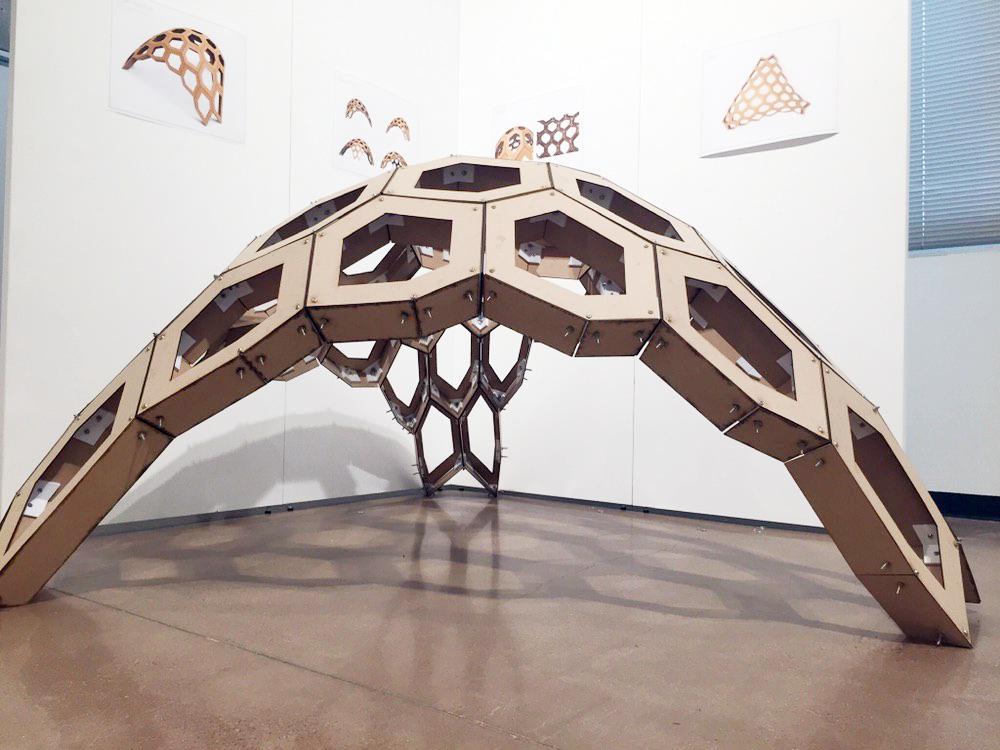}
    \caption{Completed fabricated pavilion}
    \label{fig:pavilion}
\end{figure}

The joint was defined by the pyramid $p_1 p_2 p_3 v_i$ (Fig.~\ref{fig:joint}), and thickness was added based on the stability requirements of the structure. Apertures were strategically defined on every side of the joint and its associated wall to facilitate the insertion of screws.
\begin{figure}[H]
    \centering
    \includegraphics[width=\figscale\linewidth]{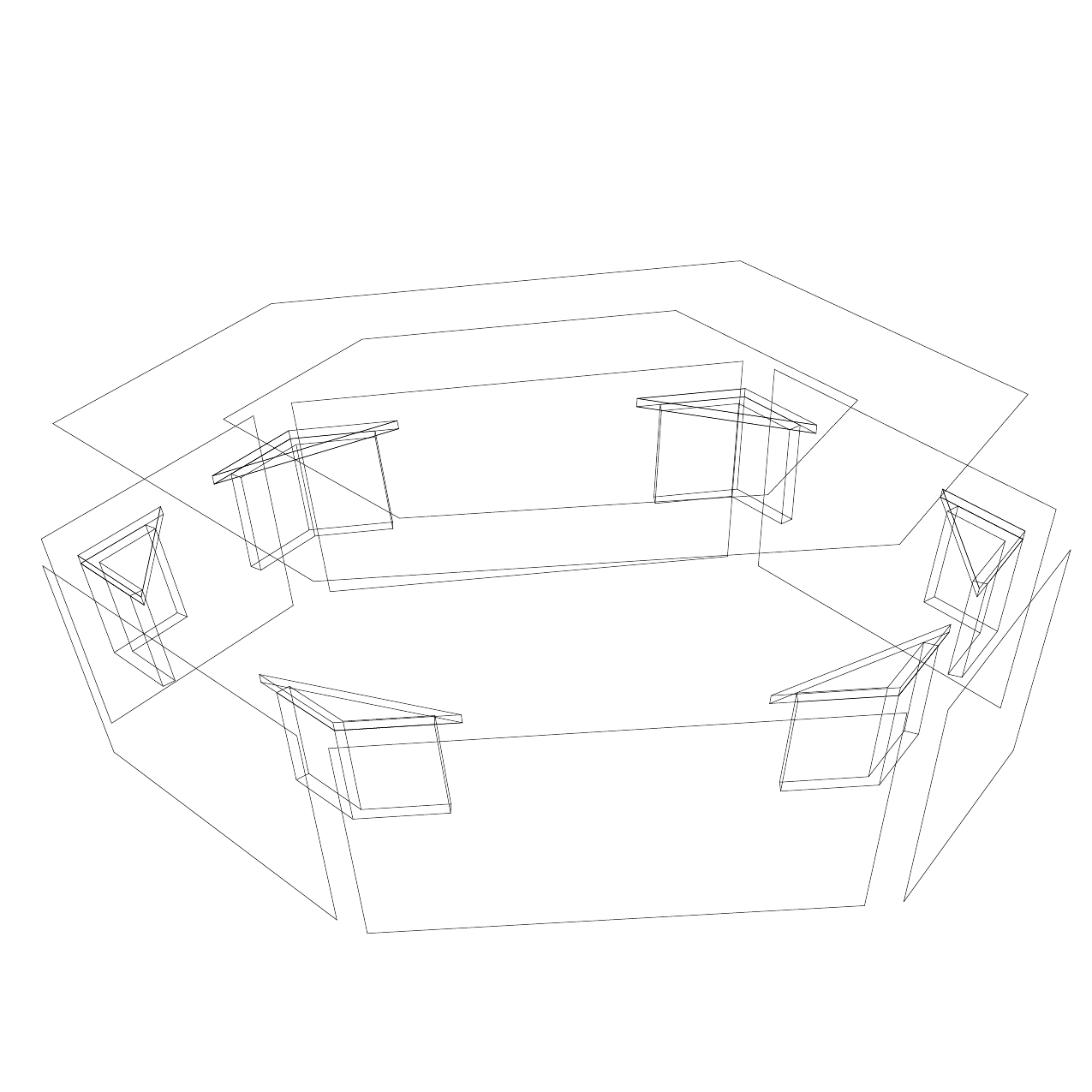}
    \caption{Panel-box joints' detail}
    \label{fig:box}
\end{figure}
Detailed views of the panel-box joints are presented in Fig.~\ref{fig:fabricated-panel}, demonstrating the physical realization of the algorithmically designed structure.
The culmination of these fabrication steps resulted in the completion of a sophisticated pavilion, as illustrated in Fig.~\ref{fig:pavilion}. We were able to successfully combine algorithmic design, physics-based simulations, and advanced fabrication techniques to make this pavilion. It shows how our method can be used to make innovative and structurally sound architectural elements.

%% file: data/06-conclusion.tex
\section{Conclusion}
\section{Conclusion}
This paper presents a comprehensive framework for computational design, centered around the Constructing Data Structure algorithm, which enables the generation of a hexagonal grid embedded within a subdivided equilateral triangle surface. This foundational step produces a robust triangular graph that supports the seamless integration of hexagons, forming the basis for downstream structural simulations.

The algorithm operates efficiently with an $O(n^2)$ complexity, highlighting its scalability for dense geometric constructions. Building upon this foundation, we introduced a Particle Spring System to model dynamic behaviors within the hexagonal mesh, providing physical realism through force-driven simulations. We further integrated a Panel Planarization optimization strategy to achieve near-planar hexagonal panels. This formulation minimizes distances between control and target points while maintaining face planarity, enabling feasible fabrication.

Our results demonstrate successful alignment between digital simulation and physical fabrication. The realized pavilion exemplifies the system's capacity to merge algorithmic precision with architectural expression. Structural reinforcement through walls and joints highlights the framework's adaptability for real-world deployment.

This approach supports a wide range of applications, from architectural pavilions to temporary structures for exhibitions and urban planning fieldwork \cite{gorjian_green_nodate, gorjian_greening_2025, gorjian_impact_2025, gorjian_schoolyard_2025}. Overall, our work bridges algorithmic design and practical construction, contributing both a theoretical advance and a validated method for creating structurally sound, visually compelling architectural systems.

%% file: ACM-PHYSICS-FROM.bbl

\begin{thebibliography}{21}

%%% the Web address.

\ifx \showCODEN    \undefined \def \showCODEN     #1{\unskip}     \fi
\ifx \showDOI      \undefined \def \showDOI       #1{#1}\fi
\ifx \showISBNx    \undefined \def \showISBNx     #1{\unskip}     \fi
\ifx \showISBNxiii \undefined \def \showISBNxiii  #1{\unskip}     \fi
\ifx \showISSN     \undefined \def \showISSN      #1{\unskip}     \fi
\ifx \showLCCN     \undefined \def \showLCCN      #1{\unskip}     \fi
\ifx \shownote     \undefined \def \shownote      #1{#1}          \fi
\ifx \showarticletitle \undefined \def \showarticletitle #1{#1}   \fi
\ifx \showURL      \undefined \def \showURL       {\relax}        \fi
% invisible to TeX
\providecommand\bibfield[2]{#2}
\providecommand\bibinfo[2]{#2}
\providecommand\natexlab[1]{#1}
\providecommand\showeprint[2][]{arXiv:#2}

\bibitem[Alexa and Wardetzky(2011)]%
        {alexa_discrete_2011}
\bibfield{author}{\bibinfo{person}{Marc Alexa} {and} \bibinfo{person}{Max
  Wardetzky}.} \bibinfo{year}{2011}\natexlab{}.
\newblock \showarticletitle{Discrete {Laplacians} on general polygonal meshes}.
\newblock In \bibinfo{booktitle}{\emph{{ACM} {SIGGRAPH} 2011 papers}}.
  \bibinfo{pages}{1--10}.
\newblock


\bibitem[Almegaard et~al\mbox{.}(2007)]%
        {almegaard_surfaces_2007}
\bibfield{author}{\bibinfo{person}{Henrik Almegaard}, \bibinfo{person}{Anne
  Bagger}, \bibinfo{person}{Jens Gravesen}, \bibinfo{person}{Bert Jüttler},
  {and} \bibinfo{person}{Zbynek Šír}.} \bibinfo{year}{2007}\natexlab{}.
\newblock \showarticletitle{Surfaces with piecewise linear support functions
  over spherical triangulations}. In \bibinfo{booktitle}{\emph{Mathematics of
  {Surfaces} {XII}: 12th {IMA} {International} {Conference}, {Sheffield}, {UK},
  {September} 4-6, 2007. {Proceedings} 12}}. \bibinfo{publisher}{Springer},
  \bibinfo{pages}{42--63}.
\newblock


\bibitem[Attar et~al\mbox{.}(2009)]%
        {attar_physics-based_2009}
\bibfield{author}{\bibinfo{person}{Ramtin Attar}, \bibinfo{person}{Robert
  Aish}, \bibinfo{person}{Jos Stam}, \bibinfo{person}{Duncan Brinsmead},
  \bibinfo{person}{Alex Tessier}, \bibinfo{person}{Michael Glueck}, {and}
  \bibinfo{person}{Azam Khan}.} \bibinfo{year}{2009}\natexlab{}.
\newblock \showarticletitle{Physics-based generative design}. In
  \bibinfo{booktitle}{\emph{{CAAD} futures conference}}.
  \bibinfo{pages}{231--244}.
\newblock


\bibitem[Baraff and Witkin(1998)]%
        {baraff_large_1998}
\bibfield{author}{\bibinfo{person}{David Baraff} {and} \bibinfo{person}{Andrew
  Witkin}.} \bibinfo{year}{1998}\natexlab{}.
\newblock \showarticletitle{Large steps in cloth simulation}. In
  \bibinfo{booktitle}{\emph{Proceedings of the 25th annual conference on
  {Computer} graphics and interactive techniques}}. \bibinfo{pages}{43--54}.
\newblock


\bibitem[Cohen-Steiner et~al\mbox{.}(2004)]%
        {cohen-steiner_variational_2004}
\bibfield{author}{\bibinfo{person}{David Cohen-Steiner},
  \bibinfo{person}{Pierre Alliez}, {and} \bibinfo{person}{Mathieu Desbrun}.}
  \bibinfo{year}{2004}\natexlab{}.
\newblock \showarticletitle{Variational shape approximation}.
\newblock In \bibinfo{booktitle}{\emph{{ACM} {SIGGRAPH} 2004 {Papers}}}.
  \bibinfo{pages}{905--914}.
\newblock


\bibitem[Cutler and Whiting(2007)]%
        {cutler_constrained_2007}
\bibfield{author}{\bibinfo{person}{Barbara Cutler} {and} \bibinfo{person}{Emily
  Whiting}.} \bibinfo{year}{2007}\natexlab{}.
\newblock \showarticletitle{Constrained planar remeshing for architecture}. In
  \bibinfo{booktitle}{\emph{Proceedings of {Graphics} {Interface} 2007}}.
  \bibinfo{pages}{11--18}.
\newblock


\bibitem[Gorjian({[n.\,d.]})]%
        {gorjian_green_nodate}
\bibfield{author}{\bibinfo{person}{Mahshid Gorjian}.}
  \bibinfo{year}{[n.\,d.]}\natexlab{}.
\newblock \showarticletitle{Green schoolyard investments influence local-level
  economic and equity outcomes through spatial-statistical modeling and
  geospatial analysis in urban contexts}.
\newblock  (\bibinfo{year}{[n.\,d.]}).
\newblock
\urldef\tempurl%
\url{https://osf.io/td9bm/download}
\showURL{%
\tempurl}
\newblock
\shownote{Publisher: OSF}.


\bibitem[Gorjian(2025a)]%
        {gorjian_greening_2025}
\bibfield{author}{\bibinfo{person}{Mahshid Gorjian}.}
  \bibinfo{year}{2025}\natexlab{a}.
\newblock \bibinfo{title}{Greening {Schoolyards} and the {Spatial}
  {Distribution} of {Property} {Values} in {Denver}, {Colorado}}.
\newblock
\newblock
\urldef\tempurl%
\url{https://doi.org/10.48550/arXiv.2507.08894}
\showDOI{\tempurl}
\newblock
\shownote{arXiv:2507.08894 [physics]}.


\bibitem[Gorjian(2025b)]%
        {gorjian_impact_2025}
\bibfield{author}{\bibinfo{person}{Mahshid Gorjian}.}
  \bibinfo{year}{2025}\natexlab{b}.
\newblock \bibinfo{title}{The {Impact} of {Greening} {Schoolyards} on
  {Residential} {Property} {Values}}.
\newblock
\newblock
\urldef\tempurl%
\url{https://osf.io/mp47r/download}
\showURL{%
\tempurl}


\bibitem[Gorjian(2025c)]%
        {gorjian_schoolyard_2025}
\bibfield{author}{\bibinfo{person}{Mahshid Gorjian}.}
  \bibinfo{year}{2025}\natexlab{c}.
\newblock \bibinfo{title}{Schoolyard {Greening}, {Child} {Health}, and
  {Neighborhood} {Change}: {A} {Comparative} {Study} of {Urban} {U}.{S}.
  {Cities}}.
\newblock
\newblock
\urldef\tempurl%
\url{https://doi.org/10.48550/arXiv.2507.08899}
\showDOI{\tempurl}
\newblock
\shownote{arXiv:2507.08899 [physics]}.


\bibitem[Hoffmann(2010)]%
        {hoffmann_local_2010}
\bibfield{author}{\bibinfo{person}{Tim Hoffmann}.}
  \bibinfo{year}{2010}\natexlab{}.
\newblock \showarticletitle{On local deformations of planar quad-meshes}. In
  \bibinfo{booktitle}{\emph{Mathematical {Software}–{ICMS} 2010: {Third}
  {International} {Congress} on {Mathematical} {Software}, {Kobe}, {Japan},
  {September} 13-17, 2010. {Proceedings} 3}}. \bibinfo{publisher}{Springer},
  \bibinfo{pages}{167--169}.
\newblock


\bibitem[Karlberg et~al\mbox{.}(2013)]%
        {karlberg_state_2013}
\bibfield{author}{\bibinfo{person}{Magnus Karlberg}, \bibinfo{person}{Magnus
  Löfstrand}, \bibinfo{person}{Stefan Sandberg}, {and}
  \bibinfo{person}{Michael Lundin}.} \bibinfo{year}{2013}\natexlab{}.
\newblock \showarticletitle{State of the art in simulation-driven design}.
\newblock \bibinfo{journal}{\emph{International Journal of Product
  Development}} \bibinfo{volume}{18}, \bibinfo{number}{1}
  (\bibinfo{year}{2013}), \bibinfo{pages}{68--87}.
\newblock
\newblock
\shownote{Publisher: Inderscience Publishers Ltd}.


\bibitem[Kilian and Ochsendorf(2005)]%
        {kilian_particle-spring_2005}
\bibfield{author}{\bibinfo{person}{Axel Kilian} {and} \bibinfo{person}{John
  Ochsendorf}.} \bibinfo{year}{2005}\natexlab{}.
\newblock \showarticletitle{Particle-spring systems for structural form
  finding}.
\newblock \bibinfo{journal}{\emph{Journal of the international association for
  shell and spatial structures}} \bibinfo{volume}{46}, \bibinfo{number}{2}
  (\bibinfo{year}{2005}), \bibinfo{pages}{77--84}.
\newblock
\newblock
\shownote{Publisher: International Association for Shell and Spatial Structures
  (IASS)}.


\bibitem[Nicolis(1993)]%
        {nicolis_physics_1993}
\bibfield{author}{\bibinfo{person}{Gregoire Nicolis}.}
  \bibinfo{year}{1993}\natexlab{}.
\newblock \showarticletitle{Physics of far-from-equilibrium systems and
  self-organization}.
\newblock In \bibinfo{booktitle}{\emph{The new physics}}.
\newblock


\bibitem[Poranne et~al\mbox{.}(2013)]%
        {poranne_interactive_2013}
\bibfield{author}{\bibinfo{person}{Roi Poranne}, \bibinfo{person}{Elena
  Ovreiu}, {and} \bibinfo{person}{Craig Gotsman}.}
  \bibinfo{year}{2013}\natexlab{}.
\newblock \showarticletitle{Interactive {Planarization} and {Optimization} of
  {3D} {Meshes}}.
\newblock \bibinfo{journal}{\emph{Computer Graphics Forum}}
  \bibinfo{volume}{32}, \bibinfo{number}{1} (\bibinfo{date}{Feb.}
  \bibinfo{year}{2013}), \bibinfo{pages}{152--163}.
\newblock
\showISSN{01677055}
\urldef\tempurl%
\url{https://doi.org/10.1111/cgf.12005}
\showDOI{\tempurl}


\bibitem[Pottmann et~al\mbox{.}(2007)]%
        {pottmann_geometry_2007}
\bibfield{author}{\bibinfo{person}{Helmut Pottmann}, \bibinfo{person}{Yang
  Liu}, \bibinfo{person}{Johannes Wallner}, \bibinfo{person}{Alexander
  Bobenko}, {and} \bibinfo{person}{Wenping Wang}.}
  \bibinfo{year}{2007}\natexlab{}.
\newblock \showarticletitle{Geometry of multi-layer freeform structures for
  architecture}.
\newblock In \bibinfo{booktitle}{\emph{{ACM} {SIGGRAPH} 2007 papers}}.
  \bibinfo{pages}{65--es}.
\newblock


\bibitem[Rezaiee-Pajand and Mohammadi-Khatami(2019)]%
        {rezaiee-pajand_fast_2019}
\bibfield{author}{\bibinfo{person}{Mohammad Rezaiee-Pajand} {and}
  \bibinfo{person}{Mohammad Mohammadi-Khatami}.}
  \bibinfo{year}{2019}\natexlab{}.
\newblock \showarticletitle{A fast and accurate dynamic relaxation scheme}.
\newblock \bibinfo{journal}{\emph{Frontiers of Structural and Civil
  Engineering}} \bibinfo{volume}{13}, \bibinfo{number}{1} (\bibinfo{date}{Feb.}
  \bibinfo{year}{2019}), \bibinfo{pages}{176--189}.
\newblock
\showISSN{2095-2449}
\urldef\tempurl%
\url{https://doi.org/10.1007/s11709-018-0486-2}
\showDOI{\tempurl}


\bibitem[Rombouts et~al\mbox{.}(2019)]%
        {rombouts_fast_2019}
\bibfield{author}{\bibinfo{person}{Jef Rombouts}, \bibinfo{person}{Geert
  Lombaert}, \bibinfo{person}{Lars De~Laet}, {and} \bibinfo{person}{Mattias
  Schevenels}.} \bibinfo{year}{2019}\natexlab{}.
\newblock \showarticletitle{A fast and accurate dynamic relaxation approach for
  form-finding and analysis of bending-active structures}.
\newblock \bibinfo{journal}{\emph{International Journal of Space Structures}}
  \bibinfo{volume}{34}, \bibinfo{number}{1-2} (\bibinfo{date}{March}
  \bibinfo{year}{2019}), \bibinfo{pages}{40--53}.
\newblock
\showISSN{0956-0599, 2059-8033}
\urldef\tempurl%
\url{https://doi.org/10.1177/0956059919864279}
\showDOI{\tempurl}


\bibitem[Schek(1974)]%
        {schek_force_1974}
\bibfield{author}{\bibinfo{person}{H.-J. Schek}.}
  \bibinfo{year}{1974}\natexlab{}.
\newblock \showarticletitle{The force density method for form finding and
  computation of general networks}.
\newblock \bibinfo{journal}{\emph{Computer Methods in Applied Mechanics and
  Engineering}} \bibinfo{volume}{3}, \bibinfo{number}{1} (\bibinfo{date}{Jan.}
  \bibinfo{year}{1974}), \bibinfo{pages}{115--134}.
\newblock
\showISSN{00457825}
\urldef\tempurl%
\url{https://doi.org/10.1016/0045-7825(74)90045-0}
\showDOI{\tempurl}


\bibitem[Schlaich and Schlaich(2000)]%
        {schlaich_lightweight_2000}
\bibfield{author}{\bibinfo{person}{Jörg Schlaich} {and} \bibinfo{person}{Mike
  Schlaich}.} \bibinfo{year}{2000}\natexlab{}.
\newblock \showarticletitle{Lightweight structures}.
\newblock \bibinfo{journal}{\emph{Widespan roof structures}}
  \bibinfo{volume}{178} (\bibinfo{year}{2000}).
\newblock
\newblock
\shownote{Publisher: Thomas Telford Publishing}.


\bibitem[Thakur et~al\mbox{.}(2009)]%
        {thakur_survey_2009}
\bibfield{author}{\bibinfo{person}{Atul Thakur}, \bibinfo{person}{Ashis~Gopal
  Banerjee}, {and} \bibinfo{person}{Satyandra~K Gupta}.}
  \bibinfo{year}{2009}\natexlab{}.
\newblock \showarticletitle{A survey of {CAD} model simplification techniques
  for physics-based simulation applications}.
\newblock \bibinfo{journal}{\emph{Computer-Aided Design}} \bibinfo{volume}{41},
  \bibinfo{number}{2} (\bibinfo{year}{2009}), \bibinfo{pages}{65--80}.
\newblock
\newblock
\shownote{Publisher: Elsevier}.


\end{thebibliography}
